\documentclass[%
 reprint,
 amsmath,amssymb,
pra,
]{revtex4-2}

\usepackage{graphicx}
\usepackage{dcolumn}
\usepackage{bm}

\usepackage{xcolor}
\begin{document}

\preprint{APS/123-QED}

\title{Characterization and Coherent Control of Spin Qubits\\ with Modulated Electron Beam and Resonator}

\author{Soheil Yasini}
 \affiliation{School of Electrical and Computer Engineering, College of Engineering, University of Tehran, Iran}
\author{Zahra Shaterzadeh-Yazdi}%
 \email{zahra.shaterzadeh@ut.ac.ir}
\affiliation{School of Engineering Science, College of Engineering, University of Tehran, Iran}%
\author{Mahmoud Mohammad Taheri}
\affiliation{School of Electrical and Computer Engineering, College of Engineering, University of Tehran, Iran}%
\date{\today}

\begin{abstract}

The coherent dynamics and control of spin qubits are essential requirements for quantum technology. A prominent challenge for coherent control of a spin qubit in a set of qubits is the destructive effect of the applied magnetic field on the coherent dynamics of neighboring qubits due to its spatial extension. We propose a novel scheme to characterize the coherent dynamics of these quantum systems and to coherently control them using a magnetic field. Our scheme consists of a resonator that encompasses the desired quantum system and a modulated electron beam that passes through the resonator in close proximity to the quantum system of interest. The dynamics of the system is obtained by solving the Lindblad master equation. To verify the reliability of our model, we tested the model on a Potassium atom, $^{41}$K and NV$^-$ center in Diamond. The results show that by properly controlling the parameters of the resonator and the electron beam, the coherence and decoherence rates of these quantum systems can be improved. Our model has the potential to be used for characterizing different types of spin-based quantum systems, and implementing quantum logic gates for quantum computation.

\end{abstract}


\maketitle

\section{Introduction}\label{sec1}
Qubits are the primary component in many areas of quantum science and technology, including quantum computation~\cite{ref1, ref2}, data encryption, safe information transmission~\cite{ref3, ref4}, and quantum metrology and imaging~\cite{ref5, ref6}. To effectively utilize qubits, it is necessary to characterize these two-level quantum systems and optimize their characteristics~\cite{ref7, ref8}. Additionally, controlling each individual qubit with minimum effect on the coherent dynamics of the neighboring ones is a necessary requirement for a wide range of applications, from investigating fundamental quantum phenomena to quantum information computation~\cite{ref9}.

Among different types of qubits, spin qubits are prominent candidates for quantum technology. A common approach for individually controlling a spin qubit is by employing a proper magnetic field. However, due to the spatial extension of the magnetic field effect and the difficulty of localizing the field in a small volume of space, individual control of these quantum systems in a set of spin qubits is a considerable challenge~\cite{ref10}.

\begin{figure}
\includegraphics[width=5.8cm]{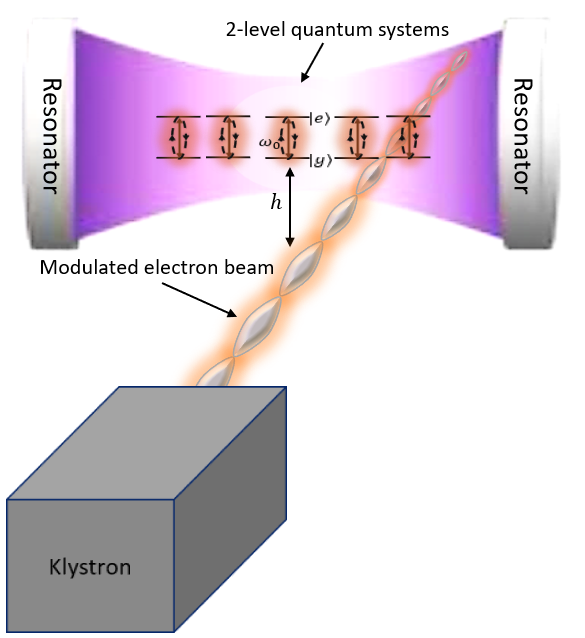}
\caption{The proposed scheme, consisting of a modulated electron beam, generated by a klystron lamp, and a resonator with a resonance frequency $\omega_r$ that is encompassing a set of two-level quantum systems with coherence frequency $\omega_0$.}
\label{fig1}
\end{figure}

In this paper, we propose a scheme for tackling this challenge, characterizing spin qubits, and controlling their coherent dynamics. As shown in Fig.~\ref{fig1}, our proposed scheme consists of a resonator encompassing a set of two-level quantum systems and a modulated electron beam that passes through a near distance $h$ from the qubit that is considered to be controlled. The desired qubit is located in the near-field region of the magnetic field generated by the electron beam. By adjusting the parameters of the resonator and the electron beam, one can characterize the dynamics of the quantum system and properly control it.

We were inspired by previous experimental and theoretical studies on measuring tunnelling rate and decoherence in double-quantum-dot systems in the microwave and terahertz regimes~\cite{ref34, ref35, ref36}. Based on these studies, the qubit of interest would resonate by using a proper frequency provided by the resonator, concomitantly with employing a proper magnetic field strength provided by the electron beam. If the frequency of the resonator is off-resonant with the coherence rate of the quantum system, it would have a small perturbative effect on the quantum system. However, if the off-resonance effect is compensated by the magnetic-field control of the electron beam, it causes resonance in the dynamics of the quantum system. The results of measurement should reveal the coherence rate and some properties of decoherence. Also, it would allow one to control over the dynamics of the quantum system by tuning the parameters of the resonator and the electron beam. 

Coherent control of an individual quantum system, with spatial resolution and on an atomic scale, becomes possible taking into account that the electron-beam characteristics are related to de-Broglie wavelength of the electrons ~\cite{ref9}, which is in the order of Angstroms and comparable to the atomic spacing. Electron-beam control of a quantum system can be achieved when the length of the wave packets associated with individual electrons is much shorter than the modulated wavelength. Such electron-beam current density modulation is easily achievable in the microwave frequency range and has wide applications in electronic technologies ~\cite{ref31}.

To investigate the dynamics of the desired quantum system, we first solve the Lindblad master equation, governing the Hamiltonian of the proposed structure and obtain the density state of the system. Quantum system’s coherence and decoherence rates are then extracted from the time evolution of the density state. In addition, we evaluate our model by considering two quantum systems, i.e.~Potassium atom ($^{41}$K) as a representative of Alkali atoms~\cite{ref32}, and Nitrogen vacancy defect centres in Diamond (NV$^-$)~\cite{ref33}. By solving the equations governing the dynamics of these quantum systems, we show that their coherence properties are improved at some frequencies of the resonator. 

This paper is organized as follows. In Sec.~\ref{sec2:background}, we provide a detailed explanation of the constituent elements of our proposed scheme. We describe the setup of the resonator and the modulated electron beam. In Sec.~\ref{sec3:method}, we introduce the model that we use in order to simulate the dynamics of the quantum system in the proposed scheme. We provide a detailed explanation of the Lindblad master equation governing the Hamiltonian of the system and how we use it to obtain the density state of the system. In Sec.~\ref{sec4:results}, we present the results of our simulations for two quantum systems, i.e.~Potassium atom and Nitrogen vacancy defect centres in diamond. Then, we discuss the implications of our results in Sec.~\ref{sec5:discussion}. Finally, in Sec.~\ref{sec6:conclusion}, we summarize the achievements of our work and suggest some future works to further develop and improve the proposed scheme.

\section{Background: Constituent Elements of the Proposed Scheme}\label{sec2:background}

\subsection{Klystron Lamp and the Electron Beam}\label{subsec2.1:klystron}

One of the key elements of the proposed scheme is the modulated electron beam. The electron beam is generated by a cathode, and then modulated by a klystron lamp. Klystron is a type of radio frequency amplifier that modulates the speed of the electron beam passing through it~\cite{ref11, ref12}. Klystrons are proposed in different dimensions and output powers, according to the need for accelerating the electrons. Also, they can operate in different operating frequencies ranging from microwave to terahertz regime~\cite{ref17, ref18, ref19}.

\begin{figure}
\includegraphics[width=9cm]{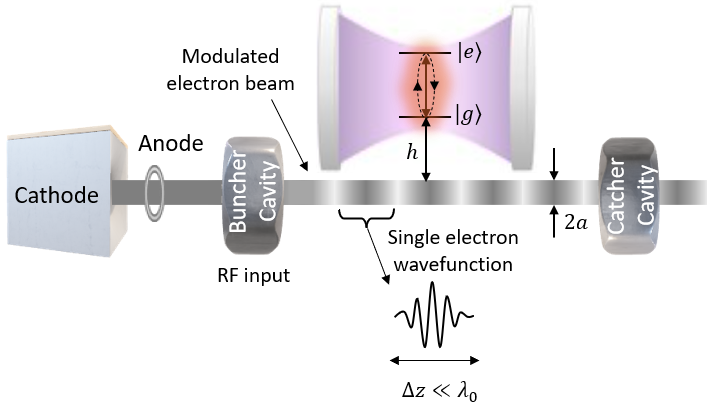}
\caption{A klystron lamp, composed of a buncher cavity and a catcher cavity. Passage of the electron beam, produced by a cathode, through the buncher cavity with frequency $\omega_0$ causes the electron beam to get modulated to a fixed modulation wavelength, $\lambda_0$. The diameter of the cross section of the electron beam is assumed to be $2a$. The wave function attributed to each electron in the electron beam is assumed to be in Gaussian form with a length $\Delta z$ much smaller than the modulation wavelength of the electron beam.}
\label{fig2}
\end{figure}

Structure of a Klystron is schematically shown in Fig.~\ref{fig2}. Klystron consists of two cavities, which are located at a certain distance from each other, along the length of the klystron. The first cavity is the buncher cavity and the second one is the catcher cavity. The electron beam is first produced by a cathode, and then moves along the length of the klystron and first passes through the buncher cavity.

Passing through the buncher cavity with frequency $\omega_0$, causes the electron beam to become modulated with a fixed modulation wavelength $\lambda_0$. The electrons are accelerated due to the applied potential difference $V_0$ between the two klystron cavities, consequently the kinetic energy of the electron beam increases. The electron beam with high kinetic energy in the output cavity produces a signal with a fixed modulation wavelength $\lambda_0$ and with a higher field amplitude. Therefore, the output signal of the klystron lamp is the same as the amplified input radiofrequency signal with a fixed modulation wavelength $\lambda_0$~\cite{ref12}.

\subsection{Resonator}\label{subsec2.2:resonator}

A resonator is usually a cavity with metal shields~\cite{ref15} or a dielectric waveguide with high permittivity~\cite{ref16}. According to its size, a resonator possesses a number of resonant frequencies that are related to different modes of electromagnetic fields. 

To model a resonator, we assume that, in its resonant-frequency mode, the resonator is equivalent to a quantum system in the excited state. Similarly, for off-resonance mode, the resonator is equivalent to the quantum system in its ground state. Therefore, a resonator with a certain resonant frequency behaves similar to a two-level quantum system, where the energy difference between its two energy levels is equivalent to the resonant frequency of the resonator.

Resonant frequency of a resonator depends on the shape, dimension, and the material of the resonator. Therefore, the value of the resonant frequency $\omega_m$ is determined, if the resonator is designed and characterized~\cite{ref13}. For a rectangular-cubic resonator with metal walls, the value for the resonant frequency of the first excited mode of the resonator, $f$, is obtained from,
\begin{eqnarray}
f=\frac{c}{2\sqrt{\epsilon_r\mu_r}}\sqrt{\left(\frac{m}{l_1}\right)^2+\left(\frac{n}{l_2}\right)^2+\left(\frac{q}{l_3}\right)^2},
\label{eqn1}
\end{eqnarray}
where $l_i$ for $i\in\{1,2,3\}$ are resonators 3D dimensions, $c$ is the speed of light in the vacuum, $\epsilon_r$ is the relative dielectric permittivity coefficient, and $\mu_r$ is the relative magnetic permeability coefficient of the material inside the cavity~\cite{ref15}. Parameters $m, n,$ and $q$ are the electromagnetic mode's arithmetic numbers, and they are not zero altogether. For higher values of these parameters, the frequency of the higher-order modes of the resonator can be obtained.

It should be noted that in resonators, made of metal shields with vacuum inside, for high frequency electromagnetic fields, the loss resulting from the dielectric property of the vacuum is small. In such cases, the electrical losses are mostly related to the currents flowing in the cavity walls. The walls are usually coated with silver or gold to increase electrical conductivity and reduce energy losses. Unlike copper, which oxidizes quickly, silver and gold are safe from oxidation~\cite{ref16}.

\section{Modelling the Proposed Scheme}\label{sec3:method}

\subsection{Basic Assumptions}\label{subsec3.1:assumptions}
To model the proposed scheme, assumptions are considered for the electron beam, the quantum system, and the resonator. As shown in Fig.~\ref{fig2}, for the electron beam, it is assumed that the wavefunction attributed to each electron is a Gaussian wave packet of length $\Delta z$ and the wavelength of the wave packet, attributed to each electron, is much smaller than the modulation wavelength of the electron beam ($\Delta z\ll \lambda_0$). It is also assumed that the electron beam current and, accordingly, the radius of the cross-section of the electron beam ($a$) is such that the distance between the center of the electron beam and the desired quantum system $h$ is large enough compared to the width of the cross-section of the beam ($2a\ll h$). 

With this assumption, the magnetic field or the non-radiative electric field caused by this modulated electron beam can be obtained from the classical relations and the interaction of the quantum system with the field caused by the electron beam can be considered semi-classical~\cite{ref9}.

Regarding the quantum system, we assume that the dimension of the desired quantum system is small compared to the wavelength of the electron beam modulation ($\lambda_0$) as well as the distance of the quantum system from the center of the electron beam ($h$). We also assume that the quantum system in question, has only two energy levels, i.e.~the ground state and the first excited state, and the probability of transition from the ground state to higher excited states is almost zero and can be ignored. 

The distance between the electron beam and the quantum system is such that the magnetic field, produced by the beam, affects the system in the near field region, thus enabling the control of the dynamics of the quantum system by the electron beam. We consider the interaction of the electron beam with the quantum system to be of a weak type, so its effect on causing decoherence in the dynamics of the quantum system can be ignored.

The resonator is assumed to be a Fabry-Perot type, and the quantum system is located inside this resonator. The resonator is used to drive the quantum system. For this purpose, the resonant frequency of the resonator is assumed to be close to the frequency associated with the energy difference between the ground and excited energy levels of the quantum system. If the frequency difference between the resonant frequency of the resonator and the quantum system is compensated by an electron beam, then the quantum system resonates and its state changes coherently between the energy levels of its ground state and excited state. 

According to the above assumptions, in the following section, we provide a suitable theoretical model to investigate the dynamics of the desired system.

\subsection{Theoretical Model for the Dynamics of the Quantum System}\label{subsec3.2:model}

We consider the quantum system and the resonator as the main system, which is affected by the electron beam. The Hamiltonian governing the whole system is defined as
\begin{eqnarray}
\hat{H}_{\text{s}}=\hat{H}_0+\hat{H}_{\text{int}},
\label{eqn2}
\end{eqnarray}
where the first term, $\hat{H}_0$, is the Hamiltonian governing the main system including the two-level quantum system and the resonator, and the second term $\hat{H}_{\text{int}}$ shows the interaction between all the components of the proposed system. 

The energy levels of the quantum system are considered to be $|g\rangle$ for the ground state and $|e\rangle$ for the excited state, and the energy difference between these two levels is equal to $\Delta E=\hbar\omega_0$. We consider the role of the resonator equivalent to quantized electromagnetic waves with the attributed energy $E_{\text{ph}}=\hbar\omega_{\text{M}}$, where $\omega_{\text{M}}$ is the resonant frequency of the resonator. Then, $\hat{H}_0$ is defined by 
\begin{eqnarray}
\hat{H}_0=\hbar \omega_0\frac{\hat \sigma_z}{2}+\hbar\omega_{\text{M}}\hat{a}_{\text{n}}^{\dagger}\hat{a}_{\text{n}}.
\label{eqn3}
\end{eqnarray}
The operator $\hat{\sigma}_z=|g\rangle\langle g|-|e\rangle\langle e|$ represents the transition between the two levels of the quantum system, and the operators $\hat{a}_{\text{n}}$ and $\hat{a}_{\text{n}}^{\dagger}$ are, respectively, the annihilation and creation operators of the quantized electromagnetic field, between the primary state $|\text{m}\rangle$ and the secondary state $|\text{m}\rangle$, attributed to the resonator. 

The second term, in Eq.~(\ref{eqn2}), is composed of the mutual interaction between the quantum system, the resonator and the electron beam and is defined by,
\begin{eqnarray}
\hat{H}_{\text{int}}=\hat{H}_{\text{int1}}+\hat{H}_{\text{int2}}+\hat{H}_{\text{int3}}.
\label{eqn4}
\end{eqnarray}
In the above equation, the first term is the interaction between the magnetic field $B$, caused by the electron beam, and the magnetic dipole moment of the two-level quantum system $\hat{\bf{\mu}}$, which is defined as~\cite{ref9},
\begin{eqnarray}
\hat{H}_{\text{int1}}=-\hat{\bf{\mu}}.\textbf{B}.
\label{eqn5}
\end{eqnarray}
We assume that the direction of the magnetic field is perpendicular to the electron beam propagation, so the mentioned interaction only happens in the $z$ direction.

The second term in Eq.~(\ref{eqn4}), $\hat{H}_{\text{int2}}$, expresses the interaction between the resonator and the quantum system, which is defined by 
\begin{eqnarray}
\hat{H}_{\text{int2}}=\gamma_{\text{n}}\left(\hat{\sigma}+\hat{\sigma}^{\dagger}\right)\left(\hat{a}_{\text{n}}+\hat{a}_{\text{n}}^{\dagger}\right),
\label{eqn6}
\end{eqnarray}
where $\hat{\sigma}=|g\rangle\langle e|$ and $\hat{\sigma}^{\dagger}=|e\rangle\langle g|$, represent the transition of the two-level quantum system from the ground state to the excited state, and from the excited state to the ground state, respectively. Parameter $\gamma_{\text{n}}$ represents the interaction coefficient between the resonator and the quantum system.  

The third term in Eq.~(\ref{eqn4}) represents the interaction between the magnetic field of the electron beam and the resonator, which is defined as,
\begin{eqnarray}
\hat{H}_{\text{int3}}=\gamma\left(\hat{a}_{\text{n}}^{\dagger}+\hat{a}_{\text{n}}\right)\left(\hat{a}_{\text{e}}^{\dagger}+\hat{a}_{\text{e}}\right)
\label{eqn7}
\end{eqnarray}
where $\gamma$ is the interaction coefficient between the magnetic field of the electron beam and the resonator. We consider a weak interaction between the magnetic field and the quantum system as well as the resonator. Therefore, we consider classical behavior for the applied magnetic field and the relations for these two parts are semi-classical.

To check the dynamics of the desired system, we consider the quantum system and the resonator together, as an open system interacting with the electron beam and assume that this interaction is weak. In order to investigate the temporal evolution of the density matrix of an open quantum system with weak interaction, we use the Lindblad equation, which is one of the well-known theories for use in this type of systems~\cite{ref21, ref22, ref23, ref24, ref25}. This equation for our desired system turns to be

\begin{eqnarray}
\frac{\partial\rho}{\partial t}=&-&\frac{i}{\hbar}\left[\hat{H}_{\text{s}},\rho\right]+\frac{\gamma}{2}\left(2\hat{\sigma}\rho\hat{\sigma}^{\dagger}-\hat{\sigma}^{\dagger}\hat{\sigma}\rho-\rho\hat{\sigma}^{\dagger}\hat{\sigma}\right)\nonumber\\
&+&\frac{\gamma_{\text{n}}}{2}\left(2\hat{a}_{\text{n}}\rho\hat{a}_{\text{n}}^{\dagger}-\hat{a}_{\text{n}}^{\dagger}\hat{a}_{\text{n}}\rho-\rho\hat{a}_{\text{n}}^{\dagger}\hat{a}_{\text{n}}\right). \label{eq8}
\end{eqnarray}
In the above formula, $\hat{H}_{\text{s}}$ is the Hamiltonian of the whole system given in Eq.~(\ref{eqn2}). Also, the density state of the quantum system is considered as $\rho=\rho_1\otimes\rho_2$, where $\rho_1$ is the density matrix of the quantum system and $\rho_2$ is the density matrix of the resonator. Taking into account the assumptions, mentioned earlier, the corresponding matrix gets the following form
\begin{eqnarray}
  \rho = 
  \begin{pmatrix}
    \rho_{\text{gg}}\rho_{\text{mm}} & \rho_{\text{gg}}\rho_{\text{mn}} & \rho_{\text{ge}}\rho_{\text{mm}} & \rho_{\text{ge}}\rho_{\text{mn}} \\
       \rho_{\text{gg}}\rho_{\text{nm}} & \rho_{\text{gg}}\rho_{\text{nn}} & \rho_{\text{ge}}\rho_{\text{nm}} & \rho_{\text{ge}}\rho_{\text{nn}} \\
   \rho_{\text{eg}}\rho_{\text{mm}} & \rho_{\text{eg}}\rho_{\text{mn}} & \rho_{\text{ee}}\rho_{\text{mm}} & \rho_{\text{ee}}\rho_{\text{mn}} \\
       \rho_{\text{eg}}\rho_{\text{nm}} & \rho_{\text{eg}}\rho_{\text{nn}} & \rho_{\text{ee}}\rho_{\text{nm}} & \rho_{\text{ee}}\rho_{\text{nn}}
  \end{pmatrix}.
  \label{eqn8}
  \end{eqnarray}
 From Eq.~(\ref{eqn8}), 16 differential equations are obtained for all the possible states of the system. Solutions of these equations expresses the time evolution of different states of the system. 
 
 By simulating the equations governing the density state of the quantum system, using finite-difference time-domain method (FDTD), and optimizing the time response of these states according to different values of the existing parameters for tuning the system, we can improve the time characteristics and the coherent dynamics of the quantum system. In the following section, we apply this model to  a couple of known quantum systems, and show the reliability of our model.
 
\section{Results}\label{sec4:results}
In this section, the results of numerical simulation for the dynamics associated with temporal evolution of the proposed system, are examined. We assume that the system is initially in one of the two states $\rho_{\text{gg}}\rho_{\text{nn}}$ or $\rho_{\text{ee}}\rho_{\text{mm}}$. These two states correspond to the cases where the quantum system is in its ground state $|g\rangle$ and the resonator is in its excited state $|n\rangle$, or vice versa, the quantum system is in the excited state $|e\rangle$ and the resonator is in its ground state $|m\rangle$. In fact, the quantum system evolves between its ground state and excited state, due to receiving and returning energy to the resonator. 

We consider two quantum systems, i.e.~Potassium atom $^{41}$K, as a representative of Alkali atoms, and NV$^-$ centers in Diamond. By using the available parameters for these two quantum systems, which are given in the following, the results obtained from the simulation are compared with the results presented in reference~\cite{ref9}.

The transition frequency between two energy levels of an individual $^{41}$K atom and similarly an NV$^-$ center in Diamond, is equal to $254$ MHz and $2.78$~GHz, respectively~\cite{ref9}. These frequencies are equivalent to $\omega_0=1.60\times 10^9$ rad/s and $\omega_0=17.34\times10^9$~rad/s, respectively. The coefficients $\gamma$ and $\gamma_n$ in Eq.~(\ref{eqn8}) are taken equal to $150$~Hz. The frequency of the resonator in transaction with the two-level quantum system is assumed to be $\omega_{\text{M}}=1.13\times10^9$~rad/s for the $^{41}$K atom, and $\omega_{\text{M}}=12.26\times 10^9$~rad/s for the NV$^-$ center.

The electron-beam current is chosen to be $100~\mu$A and the radius of the cross section of the electron beam is taken to be $25~\mu$m. The magnetic field resulting from such electron beam at the location of the quantum system is chosen to be $B=3\times10^{-9}$~T. The relationship between the magnetic field resulting from the electron beam current and the distance of the beam from the quantum system is given by the  Ampere’s law~\cite{ref20}. Hence, the distance of the beam to the quantum system, $h$, in order to obtain a magnetic field of $3\times10^{-9}$~T in the vicinity of the quantum system, is obtained to be $6.7$~mm. Alternatively, for a current of $50$~nA, in order to obtain the same amount of magnetic field, $h$ should be equal to $3.3~\mu$m.

\begin{figure}
\includegraphics[width=9cm]{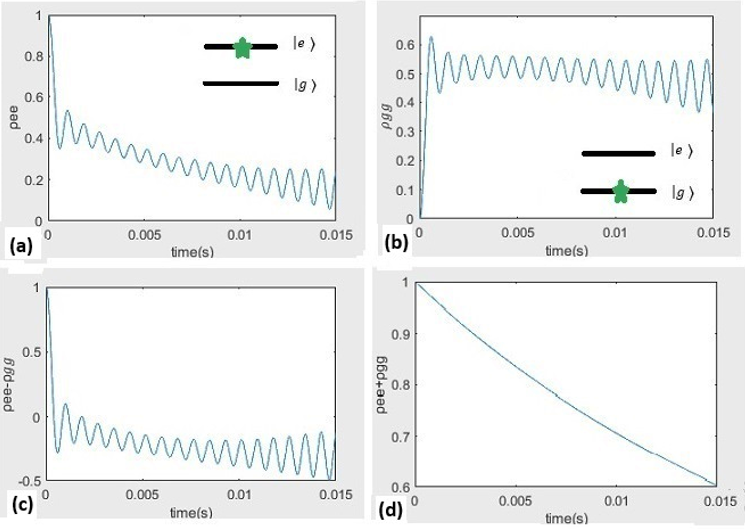}
\caption{Time evolution of the $^{41}$K atom, when (a) the atom is initially in the excited state $|e\rangle$ and (b) the atom is initially in the ground state $|g\rangle$. In the inset of each diagram, the initial quantum state of the atom is shown. (c) shows the difference between the two states and (d) demonstrates the sum of the two states.}
\label{fig3}
\end{figure}

To obtain an electron beam with the desired current value and electron-beam cross section, we need to know the voltage that is applied to the klystron cavities, in order to evaluate the energy of the electron-beam output from the klystron lamp. We assume that the electrons, with initial speed $u_0$, have a kinetic energy equal to $18,000$~eV. According to the relation between the potential difference $V_0$, between the two cavities of the Klystron, and the initial speed of the electron beam $u_0$, it is possible to estimate the value of $V_0$, necessary to produce the desired kinetic energy, based on the law of conservation of energy~\cite{ref12}. Therefore, to produce electrons with the assumed initial kinetic energy, the value of the applied voltage should be equal to $V_0=15.750$~Kv, taking into account the relativistic conditions of the speed of these particles. For these calculations, we used CST software~\cite{ref13}.

\begin{figure}
\includegraphics[width=9cm]{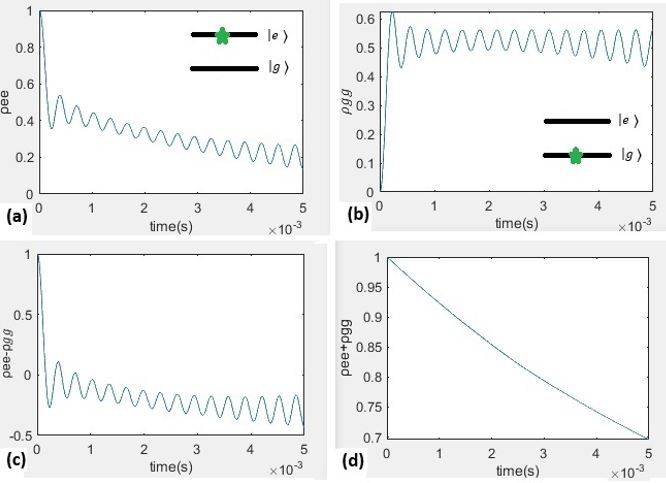}
\caption{Time evolution of the NV$^-$ center in Diamond for the case where (a) the NV$^-$ is initially in its excited state $|e\rangle$ and (b) the NV$^-$ center is initially in its ground state $|g\rangle$. In the inset of each diagram, the initial quantum state of the NV$^-$ center is shown. (c) shows the difference between the two states and (d) the sum of the two states.}
\label{fig4}
\end{figure}

In Fig.~\ref{fig3}, the time-evolution diagrams for the two states, $\rho_{\text{ee}}$ and $\rho_{\text{gg}}$, of the $^{41}$K atom are presented. The initial quantum state of the $^{41}$K atom is shown in the inset of each diagram. In Figs.~\ref{fig3}a (\ref{fig3}b), it is assumed that the atom is in the excited state $|e\rangle$ (the ground state $|g\rangle$), at the initial time $t=0$. In Figs.~\ref{fig3}c and~\ref{fig3}d, the difference and sum of the two states $\rho_{\text{ee}}$ and $\rho_{\text{gg}}$ are shown, respectively, as a function of time. Similar calculations were performed for the case of NV$^-$, where the results are presented in Fig.~\ref{fig4}. In both cases, the diagrams display valuable points about the dynamical behavior of the entire system, which will be discussed in the next section.

\section{Discussion}\label{sec5:discussion}

Comparison between Figs.~\ref{fig3}c and~\ref{fig3}d for the $^{41}$K atom, and similarly Figs.~\ref{fig4}c and~\ref{fig4}d for the NV$^-$ center, shows that for the case $\rho_{\text{ee}}-\rho_{\text{gg}}$, the time evolution of the desired quantum system has an oscillatory behavior, whereas $\rho_{\text{ee}}+\rho_{\text{gg}}$ decreases exponentially. According to the $\pm$ sign between the states $\rho_{\text{ee}}$ and $\rho_{\text{gg}}$, these results can be an indication that if the time evolution of the state of the resonator has a $\pi$-phase difference with the time evolution of the state of the quantum system, then the energy exchange between the resonator and the quantum system occurs effectively. Consequently, the resonator can control the state of the two-level quantum system and be the driving factor for the dynamics of the quantum system. 

Furthermore, the $\rho_{\text{ee}}-\rho_{\text{gg}}$ time-response diagrams for $^{41}$K atom and NV$^-$ center are compared with the corresponding diagrams presented in reference~\cite{ref9} for the same quantum systems and in the same time interval. We considered the same values for the used parameters so that the results can be compared. One can see that for both of the quantum systems, the number of Rabi oscillations in our proposed scheme has been almost doubled as compared to the number of Rabi oscillations in reference~\cite{ref9}. Hence, by employing our proposed scheme, we have been able to double the coherence rate in the quantum systems of interest (i.e. the $^{41}$K and NV$^-$ center), and significantly improve the dynamic behavior of these quantum systems. In fact, these results show the effect of the presence of the resonator and its key role in driving the quantum system of interest.

In addition, for evaluation of the decoherence rate, an exponential function is fitted to the damping of the oscillating graphs. For the results presented in the reference article~\cite{ref9}, this function is proportional to $f(t)\propto\text{e}^{-57.34t}$ for the $^{41}$K atom and $f(t)\propto\text{e}^{-923t}$ for the NV$^-$ center, whereas the exponential function fitted to our graphs is first proportional to $g(t)\propto\text{e}^{-4332t}$, in the descending part of the function, and  then it is proportional to $g(t)\propto\text{e}^{58.35t}$, in the ascending part of the function for the $^{41}$K atom. Similarly, the exponential functions fitted to the graph associated with the NV$^-$ center, in our proposed  scheme, is proportional to $g(t)\propto\text{e}^{-11770t}$, in the descending part of the function, and then it is proportional to $g(t)\propto\text{e}^{209.4t}$, in the ascending part of the function.

Comparison between the two functions $f(t)$ and $g(t)$, for each quantum system of interest, shows that the decoherence rate of the $^{41}$K atom and the NV$^-$ center in our proposed scheme, has decreased by more than $70$ times and $12$ times, respectively, as compared to the decoherence rates given in reference~\cite{ref9}. Therefore, the presence of the resonator in our proposed scheme, not only results in the increment of the coherence rate of the quantum systems, but also significantly decreases the decoherence rate. Both of these factors are important requirements for a quantum system to be recognized as a \emph{well-characterized} qubit to be used, for instance, in fault tolerant quantum computation. 

\section{Conclusion}\label{sec6:conclusion}

In this study, we proposed a scheme for characterization and coherent control of spin-based two-level quantum systems.
We showed that proper use of a resonator and a modulated electron beam lead to a better time response of the quantum states of a desired system. The modulated electron beam, generated by a Klystron lamp, is a suitable tool for individual magnetic control of a spin-based quantum system in a set of quantum systems. Also, we showed that placing a quantum system inside a proper resonator can improve the performance and characteristics of the quantum system, especially improving the coherence and decoherence rates of the system.

To model the proposed scheme, we developed a theory using the Lindblad master equation. By numerically solving the obtained equations, for the $^{41}$K atom, as a representative of Alkali atoms, and for the NV$^-$ centers in diamond, we derived the time evolution of the state of these quantum systems and examined the dynamics of the states. The results showed that specific resonance frequencies can be provided by the resonator for each quantum system, which improves the coherence dynamics of the desired system. 

Improving the coherent dynamics of a quantum system can have many applications. For instance, one can design various quantum gates for the purpose of fault tolerant quantum computation~\cite{ref26, ref27, ref28, ref29, ref30}. In fact, by controlling the tuning parameters provided by the resonator and the electron beam, suitable quantum gates can be designed and applied. Furthermore, due to the unique control that the electron beam provides for the individual access of spin-type quantum systems, it is possible to use a set of quantum systems at close proximity of each other and engineer such a compound system by electron beams for the purpose of quantum simulations and computations. Therefore, the proposed model and the equations governing its dynamics can be generalized for more than one quantum system and also for more than one resonator.

\nocite{*}

\bibliography{apssamp}

\end{document}